\documentclass{article}
\usepackage{spconf,graphicx}
\usepackage{amsmath,amssymb}

\usepackage{color}
\usepackage{float}
\usepackage{caption}
\usepackage{multirow}
\usepackage{hyperref}
\usepackage{rotating}
\usepackage{makecell}

\def\etal{{\em et al.~}}

\newcommand\blfootnote[1]{%
  \begingroup
  \renewcommand\thefootnote{}\footnote{#1}%
  \addtocounter{footnote}{-1}%
  \endgroup
}

\title{Utterance-level aggregation for speaker recognition in the wild}
\name{Weidi Xie$^{1}$, Arsha Nagrani$^{1}$, Joon Son Chung$^{1,2}$ and Andrew Zisserman$^{1}$}
\address{$^{1}$Visual Geometry Group, Department of Engineering Science, University of Oxford, UK\\
$^{2}$Naver Corporation, South Korea \\ \{weidi, arsha, joon, az\}@robots.ox.ac.uk }

\begin{document}
%\ninept
%
\maketitle
\begin{abstract}
The objective of this paper is speaker recognition `in the wild' -- where utterances may be of variable length
and also contain irrelevant signals. Crucial elements in the design of deep networks for this task are the
type of trunk (frame level) network, and the method of temporal aggregation.
We propose a powerful  speaker
recognition deep network, using a `thin-ResNet' trunk architecture,
and a dictionary-based NetVLAD or GhostVLAD layer to
aggregate features across time,  that can be trained end-to-end. We show that our network
 achieves state of the art performance by a significant margin on
the  \texttt{VoxCeleb1} test set for speaker recognition, 
whilst requiring fewer parameters than previous methods. We also investigate the effect of utterance length on performance, 
and conclude that for `in the wild' data, a longer length is beneficial.
\blfootnote{\hspace{-12pt}\url{http://www.robots.ox.ac.uk/\~vgg/research/speakerID}}

%than for clean data obtained under studio conditions.
\end{abstract}

\section{Introduction}
\label{sec:intro}
Speaker recognition `in the wild' has received an increasing amount of
interest recently due to the 
availability of free large-scale
datasets~\cite{McLaren16,Nagrani17,Chung18a}, and the 
easy accessibility of deep learning
frameworks~\cite{Abadi16,paszke2017automatic,Vedaldi15}.
For speaker recognition,
the goal is to condense information into a~\textit{single}
utterance-level representation, unlike speech recognition where
frame-level representations are desired. Obtaining a good utterance
level representation becomes particularly important for speech
obtained under noisy and unconstrained conditions, 
where irrelevant parts of the signal must be filtered out. 
Therefore, a key area of research in deep learning for speaker recognition is to investigate
how to effectively aggregate frame-level characteristics into
utterance-level speaker representations.

Earlier deep neural network (DNN) based speaker recognition systems
have na\"ively used pooling methods that have been successful for
visual recognition tasks, such as average
pooling~\cite{Nagrani17,Chung18a,li2017deep,wan2018generalized} or
fully connected layers~\cite{Lukic16,lopez2014automatic} to condense
frame-level information into utterance-level representations. Although such
methods serve the purpose of aggregating frame-level information into
a single representation whilst still allowing back-propagation, the aggregation is not
content dependent, so they are not able to 
consider which parts of the input signal contain the most
relevant information.

On the other hand, traditional methods for speaker and language
identification such as {\em i-vector} systems have explored
the use of statistical or dictionary-based methods for aggregation. A
number of recent works have proposed to bring similar methods to deep
speaker
recognition~\cite{snyder2017deep,shon2018frame,okabe2018attentive,cai2018novel,cai2018exploring} (described in Sec.~\ref{sec:related}). Based on these
works, we propose to marry the best of both Convolutional Neural Networks (henceforth, CNNs) and a
dictionary-based NetVLAD~\cite{arandjelovic2016netvlad} layer, 
where the former is known for capturing local patterns, 
and the latter can be discriminatively trained for aggregating information into a fixed-sized descriptor from an input of arbitrary size, such that the final representation of the utterance is unaffected by irrelevant information.

We make the following contributions: 
(i) We propose a powerful speaker recognition deep network, based on a NetVLAD~\cite{arandjelovic2016netvlad} or GhostVLAD~\cite{Zhong18b} layer that is used to
aggregate `thin-ResNet' architecture frame features; 
(ii) The entire network is trained end-to-end using a large margin softmax loss on the
large-scale  \texttt{VoxCeleb2}~\cite{Chung18a} dataset, 
and achieves a significant improvement over the current state-of-the-art verification
performance on  \texttt{VoxCeleb1}, 
despite using fewer parameters than the current state-of-the-art architectures~\cite{Chung18a,okabe2018attentive}; and,
(iii) We analyse the effect of input segment length on performance,
and conclude that for `in the wild' sequences having longer utterances (4s or more) is a
significant improvement over shorter segments.

\subsection{Related works}
\label{sec:related}
End-to-end deep learning based systems for speaker recognition usually
follow a similar three-stage pipeline:
 (i) frame level feature extraction using a deep neural network (DNN); (ii)
temporal aggregation of frame level features; and (iii) optimisation of a classification
loss. In the following, we review the three components in turn.

The trunk DNN architecture used is often either a 2D CNN with convolutions in both the time and frequency
domain~\cite{Nagrani17,Chung18a,cai2018exploring,cai2018analysis,hajibabaei2018unified,bhattacharya2017deep},
or a 1D CNN with convolutions applied only to the time
domain~\cite{snyder2017deep,shon2018frame,okabe2018attentive,snyder2018x}. A
number of papers~\cite{wan2018generalized, chowdhury2017attention} have
also used LSTM-based front-end architectures.

The output from the feature extractor is a variable length feature vector, 
dependant on the length of the input utterance. 
Average pooling layers have been used in~\cite{Nagrani17,Chung18a,wan2018generalized} 
to aggregate frame-level feature vectors to obtain a fixed length utterance-level embedding.
\cite{snyder2017deep} introduces an extension of the method in which the  
standard deviation is used as well as the mean 
-- this method is termed {\em statistical pooling}, 
and used by~\cite{shon2018frame,snyder2018x}.  
Unlike these methods which ingest information from all frames with equal weighting,
\cite{bhattacharya2017deep,chowdhury2017attention} have employed
attention models to assign weight to the more discriminative frames.
\cite{okabe2018attentive} combines the attention models and 
the statistical model to propose {\em attentive statistics pooling} 
-- this method holds the current state-of-the-art performance on the  \texttt{VoxCeleb1} dataset. 
The final pooling strategy of interest is the Learnable Dictionary Encoding (LDE) proposed by~\cite{cai2018novel,cai2018exploring}. 
%It is based on a dictionary learning procedure which attempts to form vectors similar to Gaussian Mixture Model supervectors~\cite{campbell2006support}. 
This method is closely based on the NetVLAD layer~\cite{arandjelovic2016netvlad, Cai_2018_ISCSLP} designed for image retrieval.

Typically, such systems are trained end-to-end for classification with a
softmax loss~\cite{okabe2018attentive} or one of its variants, such as the angular
softmax~\cite{cai2018exploring}. 
In some cases, the network is further trained for verification using the contrastive loss~\cite{Nagrani17,Chung18a,Chen11} 
or other metric learning losses such as the triplet loss~\cite{li2017deep}. 
Similarity metrics like the cosine similarity or PLDA are often adopted to generate a final pairwise score.

\section{Methods}
\label{sec:met}
For speaker recognition, the ideal model should have the following properties: 
(1) It should ingest  arbitrary time lengths as input, and produce
a fixed-length utterance-level descriptor.  
(2) The output descriptor should be {\em compact} (i.e.\  low-dimensional), 
requiring little memory, to facilitate efficient storage and retrieval.
(3) The output descriptor should also be {\em discriminative}, such that
the distance between descriptors of different
speakers is larger than those of the same speaker.

\begin{figure}[!htb]
\centering
\includegraphics[width=.48\textwidth]{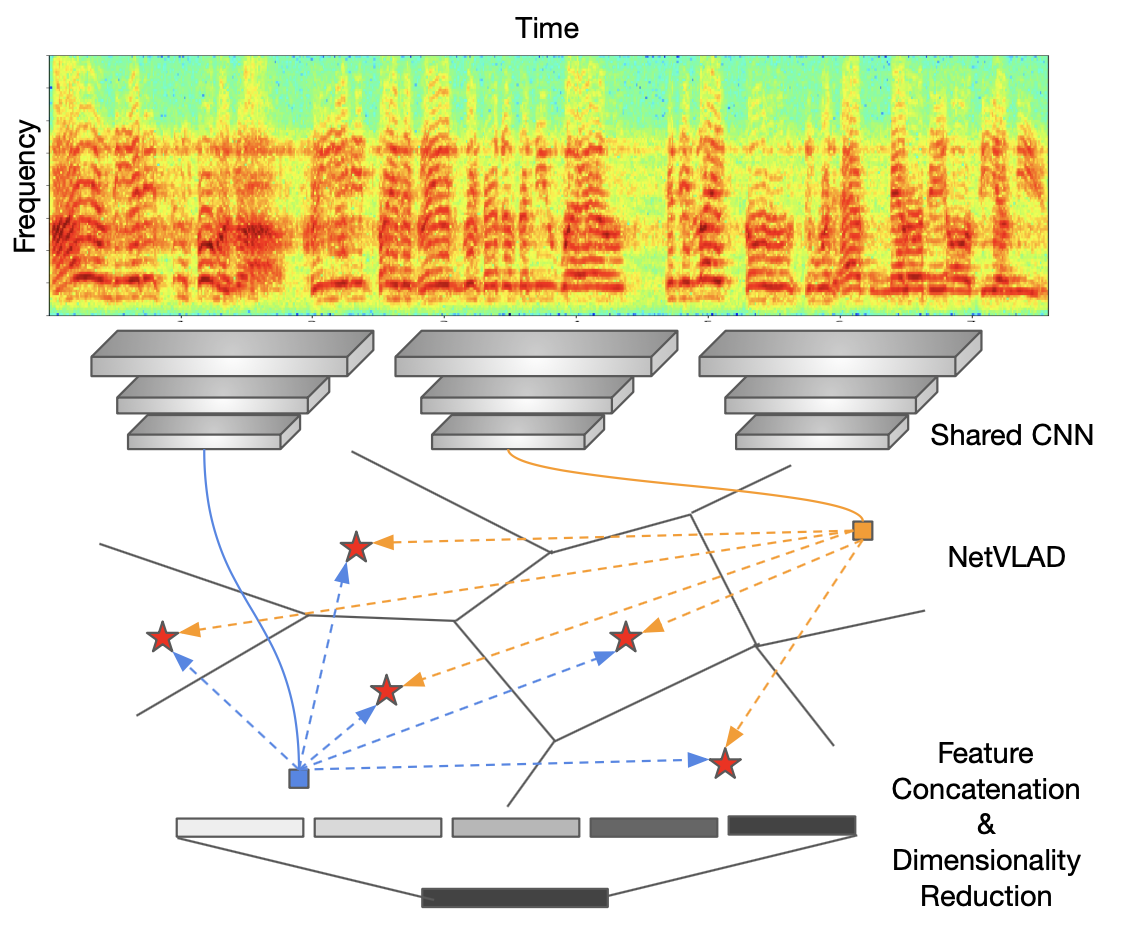}
\caption{Network architecture. It consists of two parts: 
\emph{feature extraction}, 
where a shared CNN is used to encode the spectrogram and extract frame-level features,
and \emph{aggregation}, 
which aggregates all the local descriptors into a single compact representation of arbitrary length.
}

% =====================
% 	ResNet table.
% =====================
\label{fig:arc_summary}
\end{figure}

To satisfy all the aforementioned properties, 
we use a modified ResNet in a fully convolutional way to encode input 2D spectrograms,
followed by a NetVLAD/GhostVLAD layer for feature aggregation along the temporal axis. 
This produces  a fixed-length output descriptor.
Intuitively, the VLAD layer can be thought of as trainable discriminative clustering:
every frame-level descriptor will be softly assigned to different clusters, 
and residuals are encoded as the output features. 
To allow efficient verification~(i.e.\  low memory, fast similarity computation), 
we further add a fully connected layer for dimensionality reduction.
Discriminative representations emerge because the entire network is trained end-to-end for speaker identification.
The network is shown in Figure~\ref{fig:arc_summary} and described in more detail in the following paragraphs.
% =====================
%	summary figure.
% =====================
\begin{table}[!htb]
\scriptsize
\renewcommand\arraystretch{1.4}
\begin{center}{\scalebox{0.9}{
\begin{tabular}{c|p{3.5cm}<{\centering}|p{3.5cm}<{\centering}}
% --------------------------------
\hline
Module 
& Input Spectrogram ($ 257 \times T \times 1$) 
& Output Size \\
\hline
%  ------------------
%	Embedding
%  ------------------
\multirow{16}{*}{\emph{Thin ResNet}}
& \multicolumn{1}{c|}{conv2d, $7\times7$, $64$} 
& $257 \times T \times 64$     \\
\cline{2-3}
% --------------------------------
& \multicolumn{1}{c|}{max\;pool, $2\times2$, stride ($2,2$)}
& $128 \times T/2 \times 64$     \\
\cline{2-3}
% --------------------------------
& \multicolumn{1}{c|}
{$\begin{bmatrix} {\rm conv}, 1\times 1, 48 \\ {\rm conv}, 3\times 3, 48 \\ {\rm conv}, 1\times 1, 96 \end{bmatrix} \times 2$}
& $128 \times T/2 \times 96$     \\
\cline{2-3}
% --------------------------------
& \multicolumn{1}{c|}
{$\begin{bmatrix} {\rm conv}, 1\times 1, 96\\ {\rm conv}, 3\times 3, 96\\ {\rm conv}, 1\times 1, 128 \end{bmatrix}\times 3$} 
& $ 64 \times T/4 \times 128$     \\
\cline{2-3}
% --------------------------------
& \multicolumn{1}{c|}
{$\begin{bmatrix} {\rm conv}, 1\times 1, 128\\ {\rm conv}, 3\times 3, 128\\ {\rm conv}, 1\times 1, 256 \end{bmatrix}\times 3$} 
& $ 32 \times T/8 \times 256$     \\
\cline{2-3}
% --------------------------------
& \multicolumn{1}{c|}
{$\begin{bmatrix} {\rm conv}, 1\times 1, 256\\ {\rm conv}, 3\times 3, 256\\ {\rm conv}, 1\times 1, 512 \end{bmatrix}\times 3$} 
& $ 16 \times T/16 \times 512$     \\
\cline{2-3}

& \multicolumn{1}{c|}{max\;pool, $3\times1$, stride ($2,2$)}
& $ 7 \times T/32 \times 512$    \\
\cline{2-3}

& \multicolumn{1}{c|}{conv2d, $7\times 1$, $512$} 
& $ 1 \times T/32 \times 512$    \\
\cline{1-3}

\hline
\end{tabular}}}
\end{center}
\normalsize
\vspace{-0.3cm}
\caption{The thin-ResNet used for frame level feature extraction. ReLU and batch-norm layers are not
shown. Each row specifies the \# of convolutional filters, their sizes, and the \# filters. This architecture has only 3 million parameters compared to the standard ResNet-34 (22 million).}
\label{net:resnet}
\end{table}

%--------------------------------------------------------------------------

\vspace{0.2cm}
\noindent{\bf Feature Extraction.} 
The first stage involves feature extraction from input spectrograms. 
While any network can be used in our learning framework, we opt for a modified ResNet with $34$ layers.
Compared to the standard ResNet used before by~\cite{Chung18a}, 
we cut down the number of channels in each residual block, making it a \emph{thin} ResNet-34 (Table~\ref{net:resnet}).

\vspace{0.2cm}
\noindent{\bf NetVLAD.} 
The second part of the network uses NetVLAD~\cite{arandjelovic2016netvlad} to aggregate frame-level descriptors 
into a single utterance-level vector.
Here we provide a brief overview of NetVLAD (for full details please refer to~\cite{arandjelovic2016netvlad}).

The thin ResNet maps the input spectrogram~($R^{257 \times T \times 1}$) to 
frame-level descriptors with size $R^{1\times T/32 \times 512}$.
The NetVLAD layer then takes dense descriptors as input and produces a single $K \times D$ matrix $V$, 
where $K$ refers to the number of chosen cluster, and $D$ refers to the dimensionality of each cluster.
Concretely, the matrix of descriptors $V$ is computed using the following equation:
\begin{align}
V(k,j) = \sum_{t=1}^{T/32} \frac{e^{w_k x_t+b_k}}{\sum_{k'=1}^K e^{w_{k}' x_t + b_{k'}}}(x_t(j) - c_k(j))
\end{align}
where $\{w_k\}, \{b_k\}$ and $\{c_k\}$ are trainable parameters, with $k \in [1, 2, . . . , K]$. 
The first term corresponds to the soft-assignment weight of the input vector $x_i$ for cluster $k$, 
while the second term computes the residual between the vector and the cluster centre. 
The final output is obtained by performing $L2$ normalisation and concatenation.
To keep computational and memory requirements low, 
dimensionality reduction is performed via a Fully Connected~(FC) layer, 
where we pick the output dimensionality to be $512$. 
We also experiment with the recently proposed {\bf GhostVLAD}~\cite{Zhong18b} layer, where some of the clusters are not included in the final concatenation,
and so do not contribute to the final representation,
these are referred to as `ghost clusters'~(we used \emph{two} in our implementation).
Therefore, while aggregating the frame-level features,
the contribution of the noisy and undesirable sections of a speech segment to normal VLAD clusters is effectively down-weighted, 
as most of their weights have been assigned to the `ghost cluster'. 
For further details, please see~\cite{Zhong18b}.

\section{Experiments}
\label{sec:exp}
\subsection{Datasets}
We train our model end-to-end on the \texttt{VoxCeleb2}~\cite{Chung18a} dataset (only on the `dev' partition, this contains speech from 5,994 speakers) for identification and test on the \texttt{VoxCeleb1} verification test sets~\cite{Chung18a}. Note that the
development set of \texttt{VoxCeleb2} is completely disjoint from the 
\texttt{VoxCeleb1} dataset ({\em i.e.} no speakers in common). 

\subsection{Training Loss}
Besides the standard softmax loss, we also experiment with the additive margin softmax~(AM-Softmax) classification loss~\cite{wang2018amsoftmax} during training.
This loss is known to improve verification performance by introducing a margin in the angular space. The loss is given by the following equation:

\begin{align}
L_i = -log\frac{e^{s(\cos\theta_{y_i} - m)}}{e^{s(\cos\theta_{y_i}-m)} + \sum_{j\neq{y_i}}e^{s\cos(\theta_j)} }
\label{eq:amsoftmax}
\end{align}
where $L_i$ refers to cost of assigning the sample to the correct class, 
$\theta_{y} = arccos(w^T x)$ refers to the angle between sample features~($x$) and the decision hyperplane ($w$), 
as both vectors have been L2 normalised.
The goal is therefore to minimise this angle by making $cos(\theta_{y_i}) - m$ as large as possible, where $m$ refers to the angular margin.
The hyper-parameter $s$ controls the ``temperature'' of the softmax loss, producing higher gradients to the well-separated samples (and further shrinking the intra-class variance).
We used the default values $m=0.4$ and $s = 30$~\cite{wang2018amsoftmax}.

\subsection{Training Details}
During training, we use a fixed size spectrogram corresponding to a 2.5 second 
temporal segment,  extracted randomly from each utterance.
Spectrograms are
generated in a sliding window fashion using a hamming window of width 25ms and step 10ms. 
We use a 512 point FFT, giving us 256 frequency components, which together with
the DC component of each frame gives a 
short-time Fourier transform (STFT) of size $257\times 250$ (frequency
$\times$temporal) out of every 2.5 second crop. The spectrogram 
is normalised by subtracting the mean and dividing by the standard deviation of all frequency components in a single time step.
No voice activity detection (VAD), or automatic silence removal is applied.
%noise augmentation (Gaussian noise, room impulse simulation) is applied.
We use the Adam optimiser with an initial learning rate of $1e-3$, 
and decrease the learning rate by 10 after every 36 epochs until convergence.

\section{Results}
\label{sec:res}
In this section we first compare the performance of our NetVLAD and GhostVLAD architectures trained using different losses to
the state of the art, and then investigate how performance varies with utterance length. 

\subsection{Verification on VoxCeleb1}
\begin{table*}[htb]
\footnotesize
\begin{center}
\begin{tabular}{l|l|c|c|c|c|c}
\hline
 & Front-end model & Loss & Dims & Aggregation & Training set & EER (\%)\\
\hline
\hline 
{\bf VoxCeleb1 test set} \\
\hline
Nagrani \etal \cite{Nagrani17} & I-vectors + PLDA& -- &-- &-- & VoxCeleb1 & 8.8\\
Nagrani \etal \cite{Nagrani17} & VGG-M & Softmax & 1024 & TAP & VoxCeleb1 & 10.2\\
\hline
Cai \etal \cite{cai2018exploring} & ResNet-34 & A-Softmax + PLDA & 128 & TAP &VoxCeleb1 & 4.46 \\
Cai \etal \cite{cai2018exploring} & ResNet-34 & A-Softmax + PLDA & 128 & SAP &VoxCeleb1 & 4.40 \\
Cai \etal \cite{cai2018exploring} & ResNet-34 & A-Softmax + PLDA & 128 & LDE &VoxCeleb1 & 4.48 \\
\hline
Okabe \etal \cite{okabe2018attentive} & TDNN (x-vector) & Softmax & 1500 & TAP &VoxCeleb1 & 4.70 \\
Okabe \etal \cite{okabe2018attentive} & TDNN (x-vector) & Softmax & 1500 & SAP &VoxCeleb1 & 4.19 \\
Okabe \etal \cite{okabe2018attentive} & TDNN (x-vector) & Softmax & 1500 & ASP &VoxCeleb1 & 3.85 \\
\hline
Hajibabaei \etal \cite{hajibabaei2018unified} & ResNet-20 & A-Softmax & 128 & TAP & VoxCeleb1 & 4.40 \\
Hajibabaei \etal \cite{hajibabaei2018unified} & ResNet-20 & AM-Softmax & 128 & TAP & VoxCeleb1 & 4.30 \\
\hline
Chung \etal \cite{Chung18a} & ResNet-34 & Softmax + Contrastive& 512 &TAP &VoxCeleb2 & 5.04 \\
Chung \etal \cite{Chung18a} & ResNet-50 & Softmax + Contrastive & 512 &TAP & VoxCeleb2 & 4.19 \\
\hline
{\bf Ours} & Thin ResNet-34 & Softmax& 512 &TAP & VoxCeleb2 & 10.48 \\
{\bf Ours} & Thin ResNet-34 & Softmax& 512 &NetVLAD & VoxCeleb2 & 3.57 \\
{\bf Ours} & Thin ResNet-34 & AM-Softmax & 512 &NetVLAD& VoxCeleb2 & 3.32 \\ 
{\bf Ours} & Thin ResNet-34 & Softmax& 512 &GhostVLAD & VoxCeleb2 &  \textbf{3.22} \\
{\bf Ours} & Thin ResNet-34 & AM-Softmax& 512 &GhostVLAD & VoxCeleb2 & 3.23 \\
{\bf Ours (cleaned $\dagger$) } & Thin ResNet-34 & Softmax& 512 &GhostVLAD & VoxCeleb2 &  3.24 \\
%{\bf Ours} & Thin ResNet-34 & AM-Softmax& 512 &GhostVLAD & VoxCeleb2 & TRAINING \\
\hline
\hline  
{\bf VoxCeleb1-E} \\ 
\hline 
Chung \etal \cite{Chung18a} & ResNet-50 & Softmax + Contrastive & 512 &TAP & VoxCeleb2 & 4.42 \\
%{\bf Ours} & Thin ResNet-34 & AM-Softmax & 512 & NetVLAD& VoxCeleb2 & 3.41 \\ 
{\bf Ours} & Thin ResNet-34 & Softmax & 512 & GhostVLAD& VoxCeleb2 & \textbf{3.24} \\ 
{\bf Ours (cleaned $\dagger$) } & Thin ResNet-34 & Softmax & 512 & GhostVLAD& VoxCeleb2 & \textbf{3.13} \\ 
\hline
\hline 
{\bf VoxCeleb1-H} \\ 
\hline 
Chung \etal \cite{Chung18a} & ResNet-50 & Softmax + Contrastive & 512 &TAP & VoxCeleb2 & 7.33 \\
%{\bf Ours} & Thin ResNet-34 & AM-Softmax & 512 &NetVLAD& VoxCeleb2 & 5.58 \\ 
{\bf Ours} & Thin ResNet-34 & Softmax & 512 &GhostVLAD& VoxCeleb2 & \textbf{5.17} \\ 
{\bf Ours (cleaned $\dagger$) } & Thin ResNet-34 & Softmax & 512 &GhostVLAD& VoxCeleb2 & \textbf{5.06} \\ 
\hline
\end{tabular}
\end{center}
\vspace{-0.3cm}
\caption{Results for verification on the original VoxCeleb1
test set~\cite{Nagrani17} and the extended and hard
test sets (VoxCeleb-E and VoxCeleb-H)~\cite{Chung18a}. 
\cite{cai2018exploring,okabe2018attentive,hajibabaei2018unified} do not report results on the VoxCeleb-E and VoxCeleb-H)~\cite{Chung18a} test sets. TAP: Temporal Average Pooling. SAP: Self-attentive Pooling Layer~\cite{cai2018exploring},
$\dagger$ Cleaned up versions of the test lists have been released publically. We encourage other researchers to evaluate on these lists.} 
\label{tab:voxceleb1_verification}
\normalsize
\end{table*}
\normalsize

The trained network is evaluated on three different test lists from
the \texttt{VoxCeleb1} dataset: (1) the original \texttt{VoxCeleb1}
test list with 40 speakers; (2) the extended
\texttt{VoxCeleb1-E} list that uses the entire \texttt{VoxCeleb1}
(train and test splits) for evaluation; and (3) the challenging
\texttt{VoxCeleb1-H} list where the test pairs are drawn from
identities with the same gender and nationality. In addition, we find that there are a small number of errors in the \texttt{VoxCeleb1-E} and \texttt{VoxCeleb1-H} lists, and hence we evaluate on a cleaned up version of both lists as well, which we release publically.  
The network is tested on the full length of the test segment.
We do not use any test time augmentation, which could potentially lead to slight performance gains. 

Table~\ref{tab:voxceleb1_verification} compares the performance of our
models to the current state-of-the-art on the original
\texttt{VoxCeleb1} test set. Our model outperforms all previous methods.
With standard softmax loss and a NetVLAD aggregation layer, it  outperforms the
original ResNet-based architecture~\cite{Chung18a} by a 
significant margin (EER of 3.57\%  vs 4.19\%)  whilst requiring far fewer parameters (10 vs 26 million).
By replacing the standard softmax with the additive margin softmax
(AM-Softmax), a further  performance gain is achieved (3.32\%  EER).
The GhostVLAD layer, which excludes irrelevant information from the aggregation, 
additionally makes a modest contribution to performance~(3.22\%  EER). 
On the challenging VoxCeleb1-H test set,  
we outperform the original ResNet-based architecture~\cite{Chung18a} (EER of 5.17\% vs 7.33\%),
which is  by a larger margin  than on the original VoxCeleb1 test set. 
The most similar architecture to ours  is the dictionary based method of Cai \etal~\cite{cai2018exploring}, 
which we also  outperform (EER of 3.22\%  vs 4.48\% ).  
We note that training a softmax loss based on features from temporal average pooling (TAP) 
yields extremely poor results (EER of 10.48\%).
We conjecture that the features from TAP are typically good at optimizing
the inter-class difference (i.e., separating different speakers), 
but not good at reducing the intra-class variation (i.e. making features of the same speaker compact).
Therefore, contrastive loss with online hard sample mining leads to a significant performance boost, as 
demonstrated in~\cite{Chung18a} for TAP. It is possible that this would also give a  performance boost for NetVLAD/GhostVLAD pooling.

% =====================================
\subsection{Additional experiment on GhostVLAD} 
In Table~\ref{tab:gvlad_ablation}, we study the effect of the number of clusters in the GhostVLAD layer.
Despite small differences in performance, we show that VLAD aggregation is robust to the  number of clusters and to two different loss functions.
\begin{table}[h]
\footnotesize
\begin{center}{\scalebox{0.9}{
\begin{tabular}{|c|c|c|c|c|c|}
\hline
\textbf{Loss}& \textbf{Aggregation} & \textbf{Clusters} & \textbf{G\_Clusters} & \textbf{EER (\%)} \\
\hline
Softmax & GhostVLAD & 8 & 2 & \textbf{3.22}  \\
AMSoftmax & GhostVLAD & 8 & 2  & 3.23  \\
\hline
Softmax & GhostVLAD & 10 & 2  & 3.37  \\
AMSoftmax & GhostVLAD & 10 & 2 & 3.34  \\
\hline
Softmax & GhostVLAD & 12 & 2  & 3.30  \\
\hline
Softmax & GhostVLAD & 14 & 2 & 3.31  \\
\hline
\end{tabular}}}
\end{center}
\vspace{-0.3cm}
\caption[caption]{Results for verification on the original VoxCeleb1 test set. 
All models used the same architecture~(Thin ResNet-34), and we vary the number of VLAD clusters and the loss function. }
\label{tab:gvlad_ablation}
\end{table}

\subsection{Probing verification based on length}
Table~\ref{tab:eer_length} shows the effect of the length of the test segment on speaker recognition performance. 
In order to provide a fair comparison on lengths up to 6 seconds, we
restricted the testing dataset (VoxCeleb1) to speech segments that were 6 seconds or longer (87,010 segments or 56.7\% of
the total dataset). 
To generate verification pairs, 
for each speaker in the VoxCeleb1 dataset~(1251 speakers in total),
we randomly sample 100 positive pairs and 100 negative pairs, 
resulting in 25,020 verification pairs.
During testing, 
segments of length 2s, 3s, 4s, 5s and 6s are randomly cropped from each verification pair. 
We repeat this process three times, and compute mean and standard deviation.

% =====================
%	EER vs Wav Length
% =====================
\begin{table}[htb]
\scriptsize
\begin{tabular}{l|c|c|c|c|c}
Segment \\length(s) & 2 & 3 & 4 & 5 & 6 \\
\hline 
\vspace{-0.15cm} \\
 EER & 7.97$\pm$0.06 & 5.73$\pm$0.04 & 4.70$\pm$0.02 & 4.10$\pm$0.02 & 3.39$\pm$0.02
\end{tabular}
\caption{Utterance length (in seconds) on performance.}
\label{tab:eer_length}
\end{table}
%\label{fig:eer_length}
%\end{figure}
\normalsize

As shown in Table~\ref{tab:eer_length}, 
there is indeed a strong correlation between verification performance and sequence length.  
For `in the wild' sequences, 
some of the data may be noise, silence, or speech from other speakers, 
and a single short sequence may be unlucky and consist of predominantly these irrelevant signals. 
As the temporal length increases, there is a higher chance of capturing relevant speech signals from the actual speaker.

\section{Conclusion}
\label{sec:conc}
In this paper, 
we have proposed a powerful speaker
recognition network, using a `thin-ResNet' trunk architecture,
and a dictionary-based NetVLAD and GhostVLAD layers to
aggregate features across time that can be trained end-to-end.
The network achieves state-of-the-art performance 
on the popular \texttt{VoxCeleb1} test set for speaker recognition,
whilst requiring fewer
parameters than previous methods.
We have also shown the 
effect  of utterance length on performance, 
and concluded that for `in the wild' data, 
a longer length is beneficial.\\

\noindent\textbf{Acknowledgements.} 
Funding for this research is provided by
the EPSRC Programme Grant Seebibyte EP/M013774/1. AN is supported by a Google PhD Fellowship in Machine Perception, Speech Technology and Computer Vision. 

% ===================
\bibliographystyle{IEEEbib}
\bibliography{shortstrings,vgg_local,vgg_other,mybib}

\end{document}